# Band filling and disorder effects on the normal state thermoelectric behavior in MgB$_2$

Ilaria Pallecchi [1], Marco Monni [2], Pietro Manfrinetti [1,3], Marina Putti [1,4]

[1] CNR-SPIN, c/o Università di Genova, via Dodecaneso 33, 16146 Genova, Italy
[2] *Independent contribution, Cagliari, Italy*
[3] *Dipartimento di Chimica e Chimica Industriale, Università di Genova, Via Dodecaneso 31, I-16146 Genova, Italy*
[4] *Dipartimento di Fisica, Università di Genova, Via Dodecaneso 31, I-16146 Genova, Italy*

**Abstract**
By a combined experimental and theoretical approach, we investigate normal state thermoelectric transport in MgB$_2$, as a probe of selective disorder and doping in the $\sigma$ and $\pi$ bands. We calculate the temperature dependent diffusive Seebeck coefficient $S_{diff}(T)$ with the Boltzmann equation resolved in relaxation time approximation, taking into account the scattering with phonons and impurities, the effect of renormalization and the effect doping in a rigid band approximation. We show that selective disorder has a sizeable effect on the $S_{diff}$ magnitude, as it tunes the relative contributions of $\sigma$ and $\pi$ bands. Disorder also affects the $S_{diff}$ temperature dependences, eventually yielding a linear $S_{diff}(T)$ behavior in the dirty limit. We also show that band filling has opposite effects on S, depending on which band dominates transport.
In parallel, we carry out Seebeck effect measurements on neutron-irradiated Mg$^{11}$B$_2$, and on two series of doped samples Mg$_{1-x}$Al$_x$B$_2$ and Mg(B$_{1-x}$C$_x$)$_2$. From comparison of calculated $S_{diff}(T)$ and experimental S(T) curves, we demonstrate that diffusive and phonon drag terms give comparable contributions in clean samples, but the phonon drag term is progressively suppressed with increasing disorder.
In C and Al doped samples we observe very different experimental behaviors in terms of sign, magnitude and temperature dependence. Indeed, notwithstanding the similar electron doping introduced by both substitutions, C or Al doping yields disorder which mainly affects either $\sigma$ or $\pi$ bands, respectively. With the help of our ab-initio approach, we are able to disentangle the several effects and prove that Seebeck coefficient is a very sensitive probe of the kind of disorder.

## 1. Introduction

Thermoelectric and related transport properties of metals are a source of information for their sensitivity to electronic structure, composition and carrier doping. Moreover, in contrast to the more conventional transport properties, second- and higher-order contributions in electron scattering appear to play a role in thermoelectricity, due to multiple physical mechanisms, one of which is the so called drag mechanism [1]. Drag contributions to the Seebeck effect arise from the coupling of charge carriers with boson excitations, which may be the very same excitations responsible for the pairing mechanism in superconductors. Indeed, the analysis of the phonon drag Seebeck contributions may give information on the strength of the electron-phonon coupling in conventional superconductors such as MgB$_2$ [2], in the same way as the magnon drag probes the coupling of electrons with spin waves in magnetic materials [3,4]. Furthermore, in multiband superconductors, the Seebeck effect and its temperature dependence allow to investigate the relative contributions of different bands to normal state transport in clean undoped, disordered and doped samples. In the case of MgB$_2$, the different parity and dimensionality of the $\sigma$ and $\pi$ bands- the former strongly anisotropic and the latter virtually isotropic - makes them almost independent channels of conductions [5], and the tuning of their relative contribution has a dramatic effect on superconducting properties, such as $H_{c2}(T)$ and its anisotropy [6,7] or London penetration depth and its anisotropy [8]. For this reason, chemical doping and controlled introduction of disorder in $MgB_2$ have been widely charted routes for selective tailoring of band contributions. Specifically, selective doping or selective introduction of disorder in σ and π bands are obtained by chemical substitution on the B and Mg site, respectively. The most studied substitution of the former type is $Mg(B_{1-x}C_x)_2$ [9], while of the latter type is $Mg_{1-x}Al_xB_2$ [10,11]. On the other hand, neutron irradiation is a very effective and controlled technique to induce disorder in bulk $MgB_2$, producing both point defects (vacancies and interstitials) and nanometric size defects, which affect progressively both reversible and irreversible superconducting properties [12]. It was shown that in highly irradiated samples the scattering times in the π and σ bands are comparable [13,14]. Crossover from two-band to single-band superconductivity was even observed in neutron-irradiated $MgB_2$ [15].

Numerous experimental studies on the Seebeck effect in $MgB_2$ can be found in literature. In undoped $MgB_2$, the Seebeck coefficient S is positive. The measured room temperature S values vary from 7 to 12 μV/K, with no clear trend of values within this range versus sample purity and $T_c$, as can be seen from the literature data collection reported in Table I. At low temperature just above $T_c$, S(T) exhibits an almost linear behavior [16,17,18] or a slight upward curvature in cleaner samples [2,19], while at higher temperatures, the temperature dependence tends to saturate. Thereby, the overall curve shape appears as a broad bump around above 150K, superimposed to a roughly linear temperature dependence. The most accredited qualitative interpretation of this S(T) behavior is that the measured S results from phonon-drag $S_{drag}$ and diffusive $S_{diff}$ contributions of comparable magnitude [2,19,20,21], the former associated to a broad maximum at $T \approx \Theta_D/5 - \Theta_D/4$ ($\Theta_D$ Debye temperature) and the latter associated to a roughly linear behavior. Seebeck anisotropy $S_{ab}/S_c$ measured on single crystals was found to be ≈3-4 in the whole temperature range from $T_c$ to room temperature [19].

**Table I.** Literature values of room temperature Seebeck coefficient of undoped $MgB_2$, reported together with sample characteristics, namely type of sample, residual resistivity ratio (RRR) and superconducting temperature $T_c$.

| S at room temperature (μV/K) | Type of sample | RRR | $T_c$ (K) | Ref. |
|---|---|---|---|---|
| 12.3, 11.7, 11.3 | single crystals |  | 38.1-38.8 | 19 |
| 12.2 (in-plane) | single crystal | 7 | 38.2 | 22 |
| 7 (in-plane) | single crystal | 5.1 | 38.5 | 23,24 |
| 10.6 | polycrystal | - | 38.4 | 25 |
| 10.5 | polycrystal | 2.1 | - | 23 |
| 10.3 | polycrystal | 2.7 | 39.0 | 26,27 |
| 10.3 | polycrystal | 3.0 | 38.8 | 28 |
| 8.9 | polycrystal | 2.3 | 37.0 | 29 |
| 8.7 | polycrystal | 5.4 | 38.7 | 30 |
| 8.6 | polycrystal | 3.3 | 38.3 | 17 |
| 7.9 | polycrystal | 3 | 38.0 | 2 |
| 7.7 | polycrystal | 3.5 | 38.5 | 31 |
| 7.6 | polycrystal | 4 | 39.2 | 18 |
| 7.2 | polycrystal | 3.5 | 38.7 | 32 |
| 7.1 | polycrystal | 1.8 | 38.2 | 33 |
| 6.9 | polycrystal | 3 | 38.0 | 20 |
| 6.8 | polycrystal | 4.8 | 39.8 | 34 |
| 6.4 | polycrystal | 3.2 | 38.6 | 35 |
| 5.7 | polycrystal | 2.5 | - | 23 |
| 4.1 | polycrystal | 3.5 | - | 16 |
| 3.4 | polycrystal | 3.6 | 40.4 | 32 |

Seebeck effect was also measured in doped samples. Al doping on the Mg site, $Mg_{1-x}Al_xB_2$ with x up to 1.0, was observed to yield a progressive decrease of S for x≤0.4 and eventually a sign change of S with increasing doping for x≥0.6 [20]. Be doping on the B sites was found to decrease S [25]. Non-magnetic Fe doping on the Mg site, $Mg_{1-x}Fe_xB_2$ with x up to 0.3, yielded almost negligible effect on the normal state Seebeck [28]. Analogously, almost negligible effect on dS/dT and S(300K) was found for low Mn (magnetic) doping on the Mg site, namely $Mg_{1-x}Mn_xB_2$ with x=0.07 [22], contrasting with the corresponding dramatic suppression of $T_c$ at the same doping level. C substitution in $Mg(B_{1-x}C_x)_2$ crystals with x=0.05 was observed to lower $S_{ab}$, turn $S_c$ negative and lower the anisotropy $S_{ab}/S_c$ as compared to the undoped crystal [22]. Co additions $(MgB_2)_{2-x}Co_x$ with x = 0.0, 0.1, 0.3 and 0.5 were observed to suppress significantly S and the broad S bump at ~150K, yielding and almost linear S(T) dependence for x=0.5 [34]. In all these works, qualitative analyses pointed out the role of selective doping in tuning σ and π band contributions to the Seebeck coefficient. However, a quantitative evaluation of such contributions is still lacking in literature. In this work, we develop calculations of diffusive Seebeck coefficient based on the Boltzmann equation resolved in relaxation time approximation (RTA) and carry out a direct comparison between theory and experiments in representative $MgB_2$ based systems, in order to clarify the roles of band filling, introduction of disorder in π and σ bands and phonon drag contribution in determining the normal state S(T) behavior. Specifically, we investigate the neutron-irradiated $Mg^{11}B_2$ and $Mg_{1-x}Al_xB_2$ and $Mg(B_{1-x}C_x)_2$ doped systems and we demonstrate that experimental Seebeck curves are dominated by the phonon drag S term in clean samples, but this contribution tends to vanish with increasing sample disorder. We also demonstrate that with C and Al doping, we can tune the diffusive contributions of the σ and π bands, $S_{diff\_\pi}$ and $S_{diff\_\sigma}$, introducing disorder in the σ and π band, respectively. Furthermore, we are able to observe the effect of band filling once the disorder produced by doping suppresses the drag contribution $S_{drag}$.

## 2. Theoretical methods

Our theoretical framework is the semiclassical theory of transport in metals based on the Boltzmann equation resolved in the so called relaxation time approximation [36]. The semiclassical theory leads to closed expressions for the response functions of interest, such as the electrical conductivity σ, the Seebeck coefficient **S**, the thermal conductivity κ. In the case of $MgB_2$ the two π and σ band contributions to transport properties have to be considered.

The electrical conductivity tensor σ can be expressed as:

$$\sigma = \sum_{i=\sigma,\pi} \int \widetilde{\sigma}_i(\varepsilon) \left(-\frac{df(\varepsilon-\mu)}{d\varepsilon}\right) d\varepsilon = \sum_{i=\sigma,\pi} \sigma_i \qquad (1)$$

where the index *i* enumerates the bands, μ is the chemical potential, df/dε is the derivative of the Fermi distribution function $f(\varepsilon) = \frac{1}{e^{\frac{\varepsilon-\mu}{k_BT}}+1}$, with $k_B$ Boltzmann constant, and

$$\widetilde{\sigma}_i(\varepsilon) = e^2\tau_i(\varepsilon-\mu)\int_{BZ}\frac{d\mathbf{k}}{4\pi^3}\delta(\varepsilon-\varepsilon_i(\mathbf{k}))\mathbf{v}_i(\mathbf{k})\mathbf{v}_i(\mathbf{k}) \qquad (2)$$

is the contribution to the conductivity tensor given by the electrons at the energy ε and band $i$. In Eq. (2), $e$ is the absolute value of the electron charge, $\varepsilon_i(\mathbf{k})$ are the electronic bands, $\tau_i(\varepsilon)$ is the relaxation time in the $i$-th band and $\mathbf{v}_i(\mathbf{k})\mathbf{v}_i(\mathbf{k})$ is the dyadic tensor constructed from the Fermi velocities $\mathbf{v}_i(\mathbf{k}) = \frac{1}{\hbar}\frac{d\varepsilon_i(\mathbf{k})}{d\mathbf{k}}$. The integration in Eq. (2) is extended over the entire Brillouin zone (BZ). In the adopted treatment of the RTA the relaxation times $\tau_i$ are specified functions of the wave vector $\mathbf{k}$, moreover they depend on $\mathbf{k}$ only through the energy ε, so $\tau_i(\mathbf{k})=\tau_i(\varepsilon_i(\mathbf{k}))$, this ensures the scattering times to have the same symmetry as the Fermi surface. The exact physical meaning of the energy dependence of $\tau_i$ is detailed in the following.

In the two band case the Seebeck coefficient is:

$$\mathbf{S} = -\frac{\sum_{i=\sigma,\pi}\int \tilde{\sigma}_i(\varepsilon)(\varepsilon-\mu)\left(-\frac{df(\varepsilon-\mu)}{d\varepsilon}\right)d\varepsilon}{eT\sum_{i=\sigma,\pi}\int \tilde{\sigma}_i(\varepsilon)\left(-\frac{df(\varepsilon-\mu)}{d\varepsilon}\right)d\varepsilon} = \frac{\sigma_\pi \mathbf{S}_\pi + \sigma_\sigma \mathbf{S}_\sigma}{\sigma_\pi + \sigma_\sigma} \quad (3)$$

Here T is the absolute temperature and $\mathbf{S}_\pi$ and $\mathbf{S}_\sigma$ denote the single band contributions to the Seebeck coefficient:

$$\mathbf{S}_i = -\frac{\int \tilde{\sigma}_i(\varepsilon)(\varepsilon-\mu)\left(-\frac{df(\varepsilon-\mu)}{d\varepsilon}\right)d\varepsilon}{eT\int \tilde{\sigma}_i(\varepsilon)\left(-\frac{df(\varepsilon-\mu)}{d\varepsilon}\right)d\varepsilon} \quad (4)$$

The arithmetic operations between tensors in equations like Eqs. (3) and (4) are performed elements by elements. Here, we are interested in comparing the simulation results with the measurements obtained on polycrystalline samples, which are isotropic, thereby the tensor must be averaged to give some scalar quantity. This is performed by taking the arithmetic average of the cartesian components S=($\mathbf{S}_{xx}$+$\mathbf{S}_{yy}$+$\mathbf{S}_{zz}$)/3. It is also worth pointing out that by considering the low temperature limit of Eq. (5) by means of the so called Sommerfeld expansion, that approximates the derivative of the Fermi distribution as a Dirac delta function -df/dε=δ(ε-$\varepsilon_F$), it is possible to derive the well-known Mott formula [37].

Input parameters of our semiclassical Boltzmann theory calculations are the electronic band structure and the scattering times, either by impurities (extrinsic mechanism) or by phonons (intrinsic mechanism). Electronic bands of undoped and Al doped $MgB_2$ were calculated *ab-initio* by Prof. S. Massidda and coworkers [38]. The scattering times are obtained by summing up electron-impurity and electron-phonon scattering rates, $1/\tau_0$ and $1/\tau_{e\text{-}ph}$, according to the Matthiessen rule. The energy and temperature independent scattering times from defects of σ and π carriers, $\tau_{0\sigma}$ and $\tau_{0\pi}$, can be extracted from experiments. The energy and temperature dependent electron phonon scattering times for each band $\tau_{e\text{-}ph(\sigma)}$ and $\tau_{e\text{-}ph(\pi)}$ can be evaluated as [39]:

$$\frac{1}{\tau_{e-ph(i)}(\varepsilon,T)} = \left(1+e^{-\varepsilon/k_BT}\right)\int_{-\omega_{max}}^{\omega_{max}}\frac{2\pi\alpha_i^2 F(\omega)}{(e^{\varepsilon/k_BT}-1)(1+e^{-(\varepsilon+\omega)/k_BT})}d\omega \quad (6)$$

In Eq. (6), $\alpha_i^2 F(\omega)$ are the Eliashberg functions, which were supplied by courtesy of O.V. Dolgov [40,41]. In figure 1, we show the temperature dependence of the calculated $\tau_{e\text{-}ph(\sigma)}(\varepsilon=0,T)$ and $\tau_{e\text{-}ph(\pi)}(\varepsilon=0,T)$, whose values range from ~$10^{-11}$ s at low temperature to ~$10^{-14}$ s at room temperature. In the whole temperature range, $\tau_{e\text{-}ph(\sigma)}$ is smaller than $\tau_{e\text{-}ph(\pi)}$, due to the stronger electron-phonon interaction of the σ band, yet the ratio $\tau_{e\text{-}ph(\pi)}(\varepsilon=0,T)/\tau_{e\text{-}ph(\sigma)(\pi)}(\varepsilon=0,T)$ only varies from ~2.8 to ~2 from low to room temperature, with a minimum value of ~1.8 reached at ~160K (see inset of figure 1).

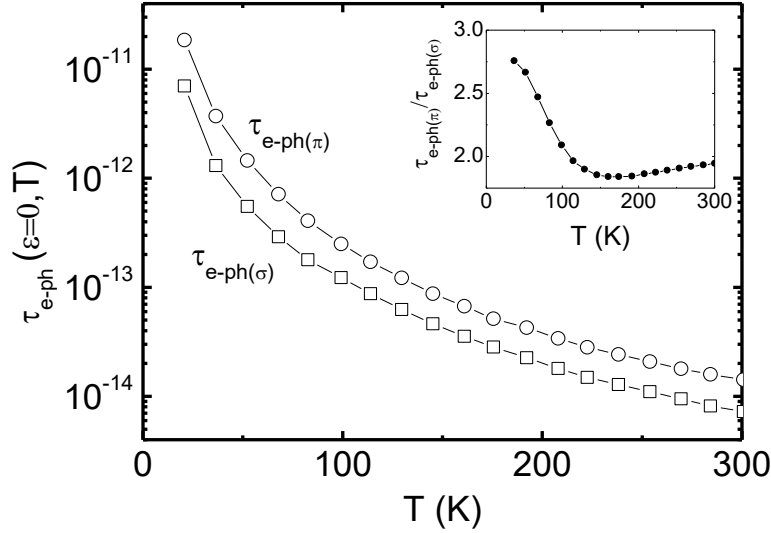

**Figure 1:** temperature dependence of the electron phonon scattering times in the two bands $\tau_{e\text{-}ph(\sigma)}(\varepsilon=0,T)$ and $\tau_{e\text{-}ph(\pi)}(\varepsilon=0,T)$; inset: ratio $\tau_{e\text{-}ph(\pi)}(\varepsilon=0,T)/\tau_{e\text{-}ph(\sigma)(\pi)}(\varepsilon=0,T)$.

With all these input parameters, the diffusive Seebeck coefficients $S_{diff}(T)$ are finally obtained by Eqs. (3) and (4).

Whether S needs some renormalization factor due to the electron phonon interaction λ or not, is a controversial point in the literature [39]. In particular A.B. Kaiser explored this possibility in detail in refs. [42,43]. Therein, he pointed out that S should be renormalized by a factor $(1+\lambda(T))$, where $\lambda(T)$ is a temperature dependent generalization of the electron phonon coupling constant λ.

## 3. Theoretical results

In order to evidence the roles of the different parameters into play, we start for simplicity by neglecting renormalization related to the temperature dependent electron-phonon coupling $\lambda(T)$. We first focus on the contributions of σ and π bands to the diffusive thermopower. We vary the level of disorder selectively in each band and study the crossover from σ to π band dominating thermoelectric transport. For simplicity, we also assume energy independent scattering times $\tau_{0\sigma}$ and $\tau_{0\pi}$ and neglect the effect of electron-phonon scattering. This limit well describes heavily doped or disordered samples in the dirty limit and yields strictly linear temperature dependence of the calculated diffusive S. In figure 2 we plot $S_{diff\_\sigma}$ and $S_{diff\_\pi}$ (thick lines): it turns out that $S_{diff\_\sigma} \gg S_{diff\_\pi}$, being $S_{diff\_\sigma} \approx 8$ μV/K and $S_{diff\_\pi} \approx 0.15$ μV/K at 300K. This is due to multiple effects, namely the lower Fermi energy of the σ band, and the hole character of the σ band as opposed to the mixed electron/hole character of the π band, with π carriers of electron and hole types that contribute oppositely to $S_{diff\_\pi}$. Given such a difference, the total Seebeck coefficient, which results from a sum of $S_{diff\_\sigma}$ and $S_{diff\_\pi}$, weighed by the respective electrical conductivities, $\sigma_\sigma$ and $\sigma_\pi$ (see eq.(3)), spans over a broad range of values, depending on the σ or π nature of disorder. We note that as long as the scattering times are independent of energy, S depends just on the ratio between the two scattering times $\eta = \tau_{\pi 0}/\tau_{\sigma 0}$ and not on the individual $\tau_{0\sigma}$ and $\tau_{0\pi}$ values. In the two panels of figure 2, we show the linear diffusive S(T) curves calculated for various values of the parameter $\eta = \tau_{0\pi}/\tau_{0\sigma}$. For η<1, the σ band dominates and for η<<1 the diffusive S attains the value $S_{diff} \approx S_{diff\_\sigma}$. For η>1, the π band dominates and diffusive S progressively decreases with increasing relative π contribution, reaching $S_{diff} \approx S_{diff\_\pi}$ for η>>1.

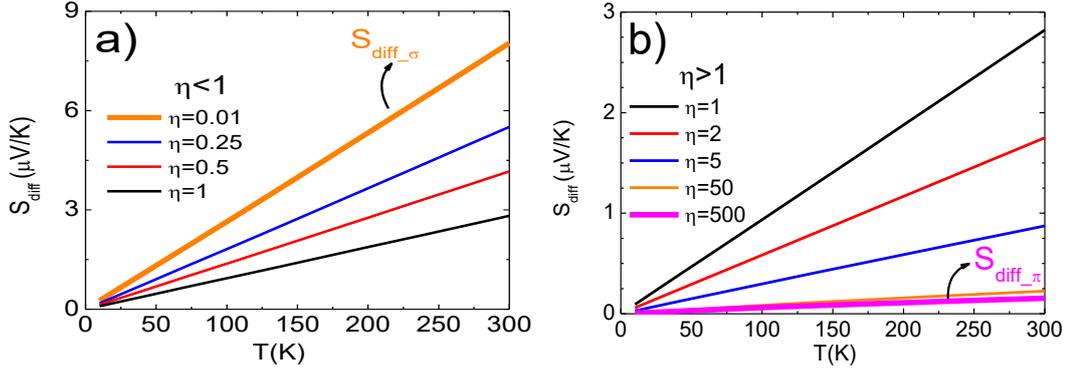

**Figure 2:** Diffusive thermopower as a function of temperature calculated in the case of energy independent scattering times and neglecting the renormalization factor, for various values of the parameter $\eta=\tau_{0\pi}/\tau_{0\sigma}$. Panel a) $\eta\leq1$, panel b) $\eta\geq1$. Thicker lines represent the limits $\eta<<1$ ($S_{diff}\approx S_{diff\_\sigma}$) and $\eta>>1$ ($S_{diff}\approx S_{diff\_\pi}$).

In order to inspect the realistic effect of disorder in each of the $\sigma$ and $\pi$ bands not only on the magnitude of $S_{diff}$, but also on the $S_{diff}(T)$ curve shape, as a result of the crossover from the clean to the dirty regime, in figure 3a and 3b we show $S_{diff\_\sigma}(T)$ and $S_{diff\_\pi}(T)$ calculated including the electron-phonon scattering time $\tau_{e-ph}$, for different values of the energy independent scattering times from impurities. In particular, in order to trace the crossover from clean to dirty limit, in panel a (b) we vary , $\tau_{0\sigma}$ ($\tau_{0\pi}$) from $10^{-9}$ to $10^{-16}$s, which largely exceeds the range of variability $\tau_{e-ph,\sigma}$ ($\tau_{e-ph,\pi}$) (see figure 1). In the clean limit ($\tau_0>>\tau_{e-ph}$), when the energy dependent electron-phonon scattering processes dominate, the magnitude of the Seebeck coefficient is lower than in the dirty limit and exhibits a bump-like feature at 75K-150K, which progressively shifts to higher temperature and flattens as $\tau_0$ decreases. The effect of disorder is that of turning the temperature dependence of the diffusive $S_{diff}$ from a bumped to a linear behavior and progressively increasing its magnitude. From the results of figures 2 and 3, we gather that the magnitude of $S_{diff}$ at room temperature reaches at most ~8.1 µV/K, in the limit $\tau_{0\pi}<<\tau_{0\sigma}<<\tau_{e-ph}$ (see figure 3a), which is smaller than most of the experimental values reported in Table I.

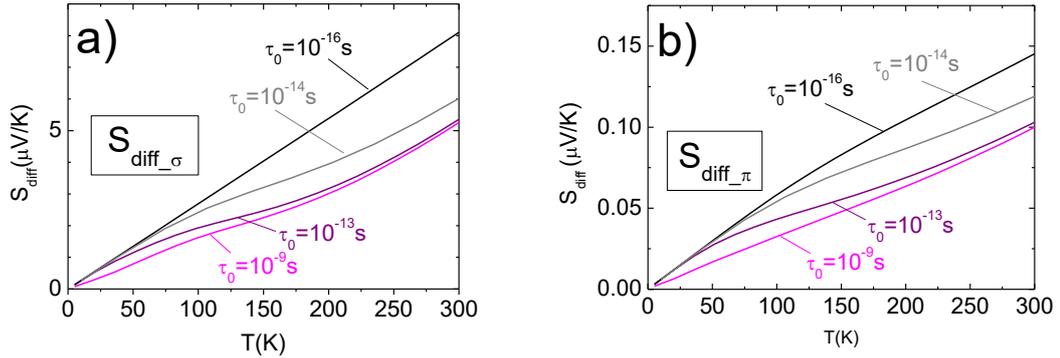

**Figure 3:** Diffusive thermopower as a function of temperature calculated including the electron-phonon scattering times, the renormalization factor and the energy independent scattering times from impurities either $\tau_{0\sigma}=\tau_0$ and $\tau_{0\pi}<<\tau_{0\sigma}$ (panel a) or $\tau_{0\pi}=\tau_0$ and $\tau_{0\sigma}<<\tau_{0\pi}$ (panel b).

In order to estimate an upper limit for the magnitude of the contributions to the diffusive S of the two bands, in figure 4 we present the calculated diffusive Seebeck coefficients of the $\sigma$ and $\pi$ bands, keeping also into account the renormalization factor $(1+\lambda(T))$, associated to the temperature dependent electron-phonon interaction $\lambda(T)$. The low temperature limit of the electron-phonon interaction is set to $\lambda(0)=0.87$, as extracted from experiments for the undoped clean MgB$_2$ [15] and in agreement with theory [10,44]. As noted above, S$_{diff\_\sigma}$>>S$_{diff\_\pi}$, indeed S$_{diff\_\sigma}$ reaches ~12 µV/K at 300K, while S$_{diff\_\pi}$ turns out to be one order of magnitude smaller ~0.22 µV/K at 300K. These curves represent an upper limit for the magnitude of the contributions to the diffusive S of the two bands and correspond to the dirty limits in both the σ and π bands, $\tau_{0\sigma}<<\tau_{0\pi},\tau_{e-ph}$ and $\tau_{0\pi}<<\tau_{0\sigma},\tau_{e-ph}$, respectively. We note that the inclusion of the renormalization factor affects also the curve shape, with a clear departure from linear temperature dependence and a progressive tendency to flatten above ~150K, indeed the renormalization factor (1+λ(T)) in MgB$_2$ reaches a maximum at intermediate temperatures ~150K [42,43].

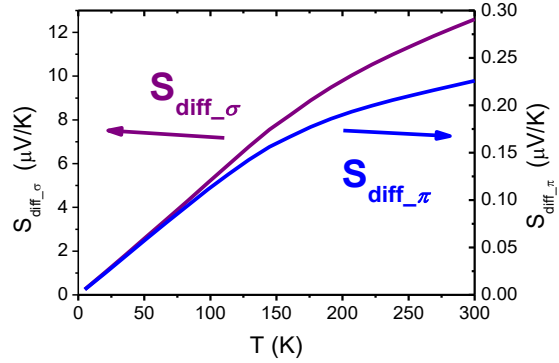

**Figure 4:** Diffusive thermopower of σ and π bands as a function of temperature calculated including the electron-phonon coupling renormalization. The low temperature limit of the electron-phonon interaction is set to the value λ(0)=0.87.

Apart from the single bands upper limits for S$_{diff}$, a band-weighed contributions should be considered for a realistic comparison with experimental data of Table I. In figure 5, we show the calculated S$_{diff}$, assuming typical values of $\tau_{0\pi}$ and $\tau_{0\sigma}$ for clean undoped MgB$_2$. In this figure, unnormalized and renormalized curves are compared. It is clear that even taking into account the renormalization, S$_{diff}$ is smaller than 4 µV/K at 300 K in this realistic case, which is well below most of the values reported in Table I. This is a clue that a missing contribution should be considered, other than the diffusive one S$_{diff}$, to account for experimental data.

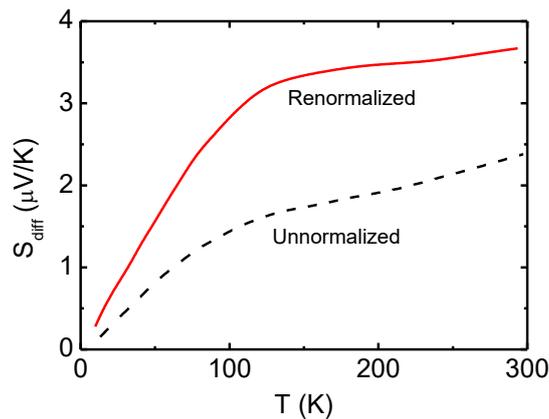

**Figure 5:** Diffusive thermopower calculated for typical values of $\tau_{0\pi}$ and $\tau_{0\sigma}$ for clean undoped MgB$_2$, $\tau_{0\pi}$=5x10$^{-15}$ s and $\tau_{0\sigma}$=2x10$^{-14}$ s. The two curves are calculated either neglecting or including the electron-phonon coupling renormalization, with the low temperature limit of the electron-phonon interaction set to the value λ(0)=0.87.

Finally, we inspect the effect of doping on the diffusive thermopower of σ and π bands, namely electron doping which is the most common in experiments. In figure 6, S$_{diff\_\sigma}$ and S$_{diff\_\pi}$ are calculated for increasing values of electron doping, in a rigid band approximation, up to *el*=0.4 electrons per unit cell. Here, we include the renormalization factor (setting $\lambda(0)=0.87$) and we consider the dirty limit, in which the scattering rates are dominated by impurity scattering ($\tau_{0\sigma}$, $\tau_{0\pi}$ << $\tau_{e-ph(\sigma)}, \tau_{e-ph(\pi)}$). It is seen that $S_{diff\_\sigma}$ is monotonically enhanced with increasing electron doping, which depletes the σ-bands and uplifts the Fermi level. Indeed, the inverse dependence of S on the carrier concentration can be explained by the Mott law, which in the low temperature limit and free electron approximation is written as $S=-(\pi^2/2e)(k_B^2 T/\varepsilon_F)$, where the Fermi energy for hole-band is estimated from the top of the band. On the other hand, $S_{diff\_\pi}$ exhibits a non-monotonic dependence on electron doping. For $el=0.1$, $S_{diff\_\pi}$ increases with respect to the undoped value, for further doping it decreases and eventually for $el=0.3$ and $el=0.4$ it turns negative, with a magnitude that increases with doping. This behavior is understood in terms of mixed hole/electron character of the π band. At low electron doping ($0 \leq el \leq 0.1$) the hole character dominates, hence the introduction of electrons increases $S_{diff\_\pi}$, as discussed above for $S_{diff\_\sigma}$. For electron doping $0.1 \leq el \leq 0.2$ holes are still dominating but the electron contribution is not negligible, hence introducing electrons is equivalent to bring the system closer to ideal electron-hole compensation, thus $S_{diff\_\pi}$ approaches zero with doping. For electron doping $0.2 \leq el \leq 0.3$ the crossover from holes to electrons occurs, and $S_{diff\_\pi}$ changes in sign. For electron doping $0.3 \leq el \leq 0.4$ electrons are dominating, thus $S_{diff\_\pi}$ is negative and increases in magnitude with doping. On the whole, the sign change of the diffusive Seebeck is a signature of electron filling in the π band (see figure 6b).

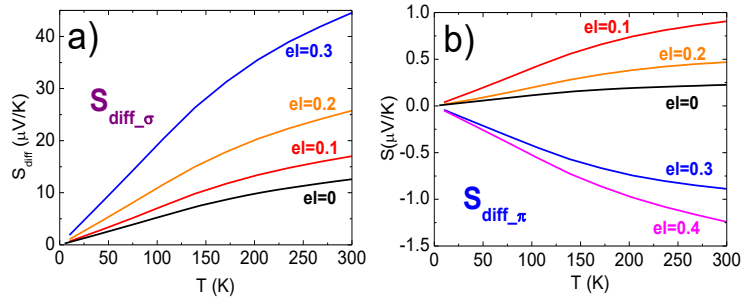

**Figure 6:** Diffusive thermopower of the σ (panel a) and π (panel b) bands as a function of temperature calculated in dirty limit ($\tau_{0\sigma}$, $\tau_{0\pi}$ << $\tau_{e-ph(\sigma)}, \tau_{e-ph(\pi)}$), including the renormalization factor for increasing values of electron doping per unit cell.

In summary, our calculations show that in undoped clean $MgB_2$ the diffusive term $S_{diff}$ is dominated by the σ band contribution, which has lower Fermi energy and higher effective mass as compared to the π band [20,21], hence the positive sign of S is accounted for in terms of σ holes. The mixed electron/hole character of the π band, which entails mutually opposite contributions to S from antibonding π holes and bonding π electrons, also contributes to the much lower S the π band as compared to the σ band [19]. The calculated $S_{diff}$ is linear only in the simplistic approximation of energy independent scattering time $\tau(\varepsilon) \approx$ constant; on the contrary, inclusion of energy dependent electron-phonon scattering time $\tau_{e-ph}(\varepsilon)$ yields departure of $S_{diff}(T)$ from linearity and appearance of bump-like features at 75K-150K (see figure 3). Taking into account the renormalization factor $(1+\lambda(T))$ [42,43], yields a broad bump at temperatures above ~150K in $S_{diff}(T)$ (see figure 5). In any case the magnitude of $S_{diff}(T)$ is well below the measured values in undoped $MgB_2$ for any realistic choice of scattering parameters.

To account for this discrepancy it is necessary to consider the phonon drag contribution. Several experimental works have indeed claimed the presence of phonon drag contribution on the basis of the temperature dependence of S curve [2,19,20,21]. In the presence of a temperature gradient, the phonons are not in thermal equilibrium and there is a net phonon flux from the hot to the cold side of the sample. If the phonons interact strongly with electrons, they transfer their momentum to the electron system and this mechanism gives an additive contribution to S, the so called phonon drag Seebeck coefficient $S_{drag}$. This mechanism is significant as long as the electron-phonon coupling is sizeable, as it is the case of conventional superconductors. A quantitative modelling of $S_{drag}$ is out of the scope of this work. A phenomenological approach in term of physical parameters is reported in ref. [21]. Here we just summarize some key qualitative aspects. At low temperature $S_{drag}$ is characterized by a $\sim T^3$ power law dependence related to the progressive thermal activation of phonon modes. With increasing temperature, approaching the Debye temperature ($\theta_D \sim 700\text{-}1000K$, the specific value depending on the experimental probe [2,45]), the phonons thermalize by themselves through umklapp processes and the drag mechanism is suppressed. As a consequence, $S_{drag}$ exhibits a broad peak around $\theta_D/5\text{-}\theta_D/4 \sim 150\text{-}250$ K. Disorder has the effect of phonon mode thermalizing and consequent suppression of $S_{drag}$.

## 4. Experimental methods
Polycrystalline samples, having stoichiometric compositions $Mg^{11}B_2$, $Mg_{1-x}Al_xB_2$ (x=0.0, 0.05, 0.10, 0.20) [11] and $Mg(B_{1-x}C_x)_2$ (x=0.0, 0.025, 0.050, 0.075, 0.100) were prepared by using a direct one-step synthesis process apt to prepare bulk samples [46,47]. Materials used are high-purity elements: Mg (99.99 wt.%), Al (99.999 wt.%), B [either 99.9 wt.% crystalline powder, or boron isotopically enriched in $^{11}$B (99.96% purity, with a residual $^{10}$B concentration lower than 0.5%)] and $B_4C$. All materials are commercial products. A master Al-Mg alloy, or $B_4C$ are used for the Al doped and the C doped samples, respectively. The synthesis is performed into outgassed Ta crucibles, sealed by arc welding in pure argon and closed under vacuum in quartz tubes, by a first heating step up to 850 °C, followed by a subsequent annealing at 950 °C for 3-4 days. Isotopically pure $Mg^{11}B_2$ polycrystals were irradiated with increasing fluence from $10^{17}$ to $10^{20}$ cm$^{-2}$ at the Paul Sherrer Institut (PSI) facility in Villigen (see details on sample preparation and irradiation in ref. [48]).
The Seebeck effect measurements are performed in a home built cryostat operating from 4K to 300K. The thermopower is measured in a steady heat flux configuration using an ac technique with frequencies 0.005-0.003 Hz. The temperature gradient, varying from 0.01 to 0.001 K/m, is measured with two Au(Fe)-Kp thermocouples. Resistivity characterization is carried out in a commercial Physical Properties Measurement System (PPMS) by Quantum Design.

## 5. Experimental results and comparison with calculated diffusive Seebeck coefficient $S_{diff}$
### 5.1 Irradiated samples
In Figure 7 the resistivity curves $\rho(T)$ and Seebeck effect curves S(T) of the neutron irradiated samples are displayed. These samples were characterized in a previous work [48]. The residual resistivity $\rho(T=40K)$ increases by two orders of magnitude with irradiation and $T_c$ decreases down to 9 K for fluence $\Phi=1.4\times10^{20}$ cm$^{-2}$ (see Table II), but the transition width remains extremely narrow, as in the pristine sample. Focusing on the Seebeck effect, the behavior of the pristine ($\Phi=0$) sample complies with expectations from literature data, exhibiting an almost linear temperature dependence just above $T_c$ and a tendency to saturation at higher temperatures, with a room temperature value $\approx 12$ μV/K. With increasing neutron fluence, S is suppressed in magnitude and its room temperature value is $\approx 3$ μV/K for fluences $\Phi \geq 2.0\times10^{18}$ cm$^{-2}$. The temperature dependence becomes more linear in the whole temperature range after irradiation. The superconducting transition temperature detected from S curves is $T_c \approx 36K$ for $\Phi=7.6\times10^{17}$ cm$^{-2}$, $T_c \approx 32K$ for $\Phi=2.0\times10^{18}$ cm$^{-2}$ $T_c \approx 8K$ for $\Phi=1.4\times10^{20}$ cm$^{-2}$, all values slightly lower but compatible with those extracted from resistivity using the 90% criterion.

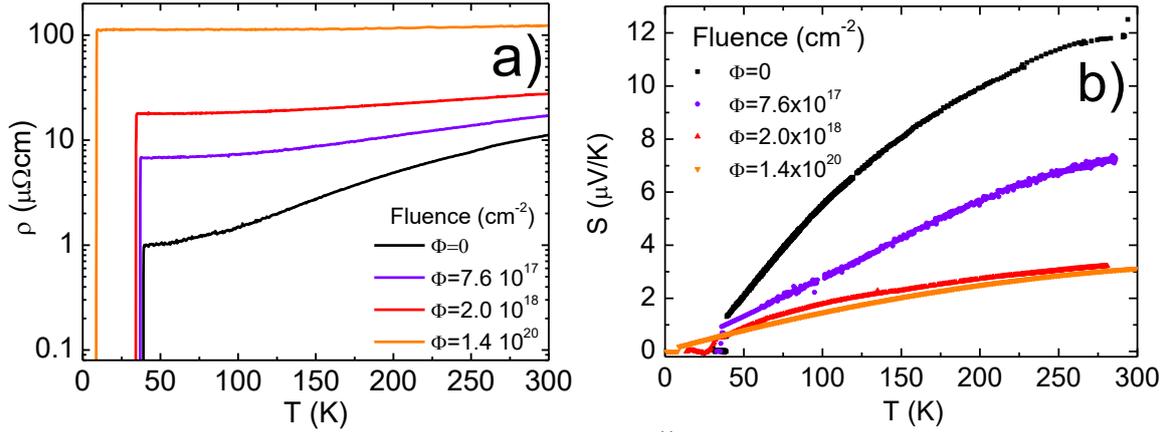

**Figure 7:** ρ(T) (panel a) and S(T) (panel b) curves measured on Mg$^{11}$B$_2$ polycrystals irradiated with increasing fluences.

We now calculate the diffusive term $S_{diff}(T)$ specifically for the analyzed samples, by fixing the free parameters $\tau_{0\sigma}$, $\tau_{0\pi}$ and $\lambda$ from experiments carried out on the very same samples. In the case of irradiated samples, $\tau_{0\sigma}$ and $\tau_{0\pi}$ values are fixed from magnetoresistance experiments [13] and the low temperature limits of the electron-phonon coupling constants $\lambda$ are estimated from specific heat measurements of the Sommerfeld constant $\gamma$ [15], which is related to $\lambda$ by means of the relation $\gamma=(2/3)\pi^2 k_B^2 N(\varepsilon_F)(1+\lambda)$, where $N(\varepsilon_F)$ the calculated density of states at the Fermi level. In Table II, the values of these parameters are listed, together with the residual resistivities and Tc's of each sample. In figure 8, we show the so calculated $S_{diff}(T)$ curves to be compared with experimental S(T) curves displayed in figure 7. It is seen that the pristine sample, which is in the clean regime, exhibits a bump at 75 K, associated to the electron-phonon scattering (see figure 3), which progressively shift to higher temperature with increasing irradiation. The most irradiated sample exhibits an almost linear temperature dependence, associated to the approaching of the dirty limit (see figure 3) and to the reduction of the renormalization factor $(1+\lambda(0))$ (see $\lambda(0)$ values in Table II). The evolution of the overall slope with increasing disorder (increasing neutron fluence) is a consequence of the crossover from σ-band dominated transport ($\eta=0.25$) to balanced contributions of both bands ($\eta=1$). The magnitude of $S_{diff}$ is intermediate between typical $S_{diff\_\sigma}$ and $S_{diff\_\pi}$ values, consistent with the values of $\eta=\tau_{0\pi}/\tau_{0\sigma}$ reported in Table II, not much departing from unity, which indicate that irradiation introduces disorder in both σ and π bands. The calculated $S_{diff}$ curves depart severely from the experiment, both in terms of curve shape and magnitude, in the cases of the clean and low irradiated samples, but they are in satisfactory agreement in the cases of the two most irradiated samples.

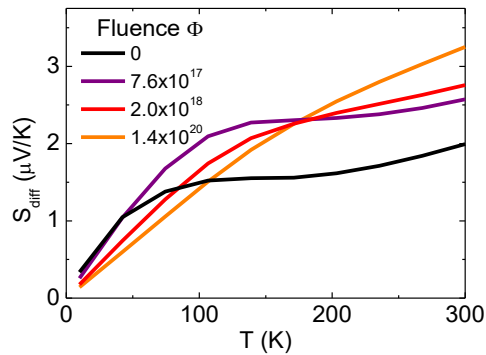

**Figure 8:** Diffusive contribution to thermopower calculated for irradiated samples, using the experimental parameters listed in Table II. The corresponding experimental S(T) curves of the same samples are displayed in figure 7b.

**Table II.** Parameters $\tau_{0\sigma}$, $\tau_{0\pi}$, $\eta=\tau_{0\pi}/\tau_{0\sigma}$ and $\lambda$ obtained from experiments carried out on the neutron irradiated samples and used for the calculation of $S_{diff}$ in these samples (figure 8). Experimental values of residual resistivity and superconducting temperature $T_c$ are also reported.

| Fluence $\Phi$ (cm$^{-2}$) | $\tau_{0\sigma}$ (s) | $\tau_{0\pi}$ (s) | $\eta$ | $\lambda$ | $\rho$ at 40K ($\mu\Omega$cm) | $T_c$ (K) |
|---|---|---|---|---|---|---|
| 0 | 2.7x10$^{-14}$ | 6.7x10$^{-14}$ | 0.25 | 0.87 | 1 | 39.2 |
| 7.6x10$^{17}$ | 2.8x10$^{-14}$ | 1.4x10$^{-14}$ | 0.5 | 0.87 | 10 | 36.1 |
| 2.0x10$^{18}$ | 6.8x10$^{-15}$ | 6.8x10$^{-15}$ | 1 | 0.87 | 17 | 33.5 |
| 1.4x10$^{20}$ | 1.0x10$^{-15}$ | 1.0x10$^{-15}$ | 1 | 0.49 | 95 | 9.2 |

## 5.2 Al doped samples

In Figure 9, we present resistivity curves $\rho(T)$ and Seebeck effect curves S(T) measured on the series of Al doped polycrystals Mg$_{1-x}$Al$_x$B$_2$. Also these samples were characterized in a previous work [11]. It is found that with increasing x the residual resistivity $\rho(T=40K)$ increases, $T_c$ decreases down to 29 K (see Table III) and the transition width remains rather narrow. With increasing Al doping, both the low temperature linear slope and the room temperature S values are weakly decreased, the latter reaching ≈9 μV/K for x=0.2 as compared to ≈11.5 μV/K for x=0. However, the overall curve shape remains virtually unchanged upon Al doping. A similar trend was found in ref. [20], where further doping values up to x=1.0 were investigated and a sign change in S was observed for x≥0.6. The superconducting transition is visible in the S(T) curves of all the samples, although with increasing x its onset decreases in temperature and its width increases ($T_c$≈39K for x=0, $T_c$≈36K for x=0.05, $T_c$≈32K for x=0.10, $T_c$≈30K for x=0.20, all values slightly lower but compatible with those extracted from resistivity using the 90% criterion).

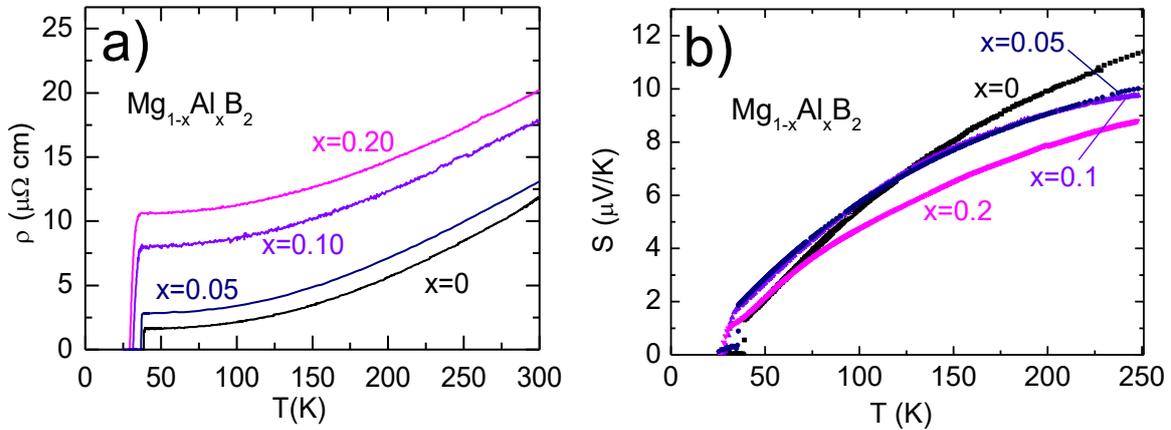

**Figure 9:** $\rho(T)$ (panel a) and S(T) (panel b) curves measured on Mg$_{1-x}$Al$_x$B$_2$ polycrystals

We now calculate diffusive $S_{diff}(T)$ for the series of Al doped samples, fixing the free parameters $\tau_{0\sigma}$ and $\tau_{0\pi}$ from experimental data of $H_{c2}$ [11] and far-infrared reflectivity [49] and $\lambda$ values from *ab initio* calculations [38]. Furthermore, the shift of the chemical potential in a rigid band approximation and the changes in the density of states at the Fermi level N($\varepsilon_F$) as a function of doping are also taken into account, as reported in the work by the group of Prof. Massidda [38]. Table III summarizes the parameters and figure 10 the corresponding simulation results. The calculated $S_{diff}$ at fixed temperature is strongly enhanced with increasing Al doping. The main reason is the σ band filling which raises the σ band contribution (see figure 6a). Indeed, in the Al doped series of samples, the σ band is the dominant one, because substitutional disorder on the Mg site does not affect scattering in σ bands, but it strongly affects π bands, as confirmed by the η values in Table III, all significantly smaller than unity and progressively decreasing with increasing Al doping. Comparing calculated and experimental data of figures 10 and 9, respectively, it emerges that there is poor agreement for the undoped and low doped samples, but a qualitative agreement can be found for the most substituted x=0.2 sample. Moreover, the trend of S as a function of Al doping is reversed, indeed the calculated $S_{diff}$ increases with doping whereas the measured S slightly decreases with doping. The observed slight decrease of S with doping cannot be ascribed to the diffusive term, as the magnitude of S should instead increase with decreasing η (see figure 2). Hence, the explanation for the discrepancy between the experimental S(T) curves and the calculated $S_{diff}$ should be found elsewhere.

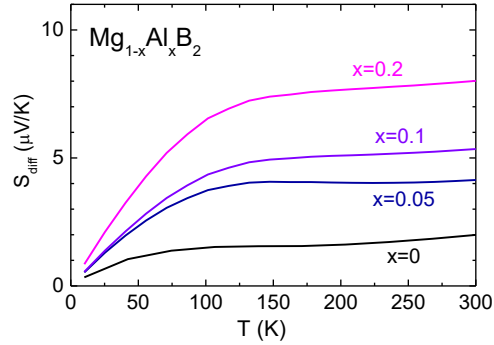

**Figure 10:** Diffusive contribution to thermopower calculated for Al doped samples, using the experimental parameters listed in Table III. The corresponding experimental S(T) curves of the same samples are displayed in figure 9b.

**Table III.** Parameters $\tau_{0\sigma}$, $\tau_{0\pi}$, $\eta=\tau_{0\pi}/\tau_{0\sigma}$ and $\lambda$ obtained from experiments and calculations carried out on the Al doped samples and used for the calculation of $S_{diff}$ in these samples (figure 10). Experimental values of residual resistivity and superconducting temperature $T_c$ are also reported.

| x doping in $Mg_{1-x}Al_xB_2$ | $\tau_{0\sigma}$ (s) | $\tau_{0\pi}$ (s) | η | λ | ρ at 40K (μΩcm) | $T_c$ (K) |
|---|---|---|---|---|---|---|
| 0 | 2.7x10$^{-14}$ | 6.7x10$^{-14}$ | 0.25 | 0.87 | 1 | 39.0 |
| 0.05 | 9.0x10$^{-14}$ | 0.9x10$^{-14}$ | 0.1 | 0.86 | 3 | 36.6 |
| 0.1 | 4.8x10$^{-14}$ | 3.8x10$^{-15}$ | 0.08 | 0.85 | 8 | 33.4 |
| 0.2 | 4.3x10$^{-14}$ | 1.7x10$^{-15}$ | 0.04 | 0.71 | 12 | 29.1 |

## 5.3 C doped samples

In Figure 11, we present resistivity curves ρ(T) and Seebeck effect curves S(T) measured on the series of C doped polycrystals $Mg(B_{1-x}C_x)_2$. The residual resistivity ρ(T=40K) increases monotonically with x, and tends to saturate for the most doped samples (x=0.075 and x=0.10), while $T_c$ decreases and the transition width increases up to 10 K. Turning to the thermoelectric behavior, S is suppressed significantly in magnitude with increasing doping and for x≥0.075 it even changes in sign and becomes negative in the whole temperature range, reaching a room temperature value ≈-4 μV/K for x=0.10. Our data are in good agreement with those measured along the ab plane in the x=0.05 C-doped single crystal of ref. [22] and with the room temperature values of $S_{ab}$ in single crystals with increasing x of ref. [50]. The curve shape is modified by doping; indeed S(T) exhibit an almost linear temperature dependence up to room temperature as long as it remains positive (0.025≤x≤0.05), while for x≥0.075 S(T) is rather flat and negative in the temperature range between 80K and 250K and increases more sharply in magnitude above and below this range. The superconducting transition is seen in the x=0.025 sample at $T_c$≈37K and at $T_c$≈25K for the highest doping.

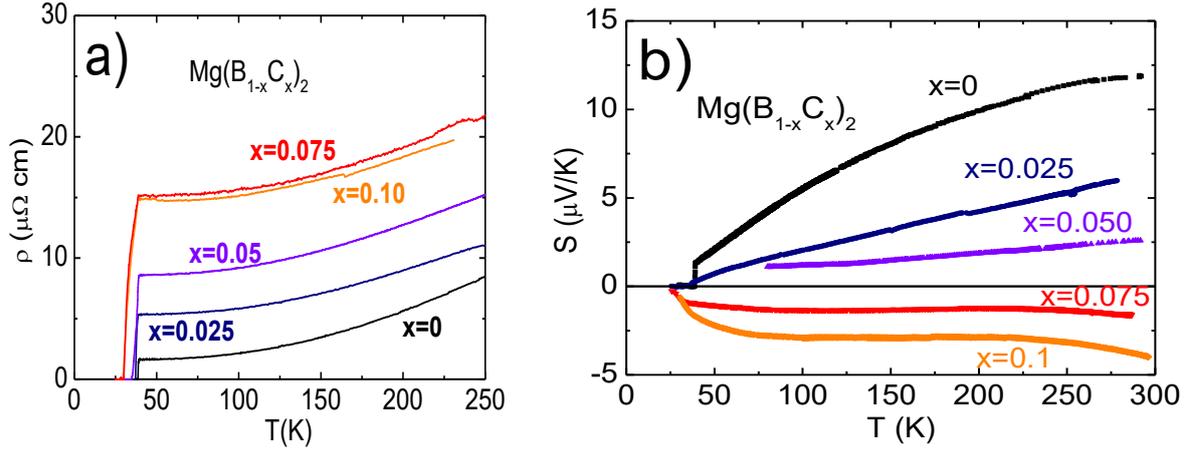

**Figure 11:** ρ(T) (panel a) and S(T) (panel b) curves measured on Mg(B$_{1-x}$C$_x$)$_2$ polycrystals

We cannot carry out quantitative simulations for the diffusive term S$_{diff}$(T) of the C doped samples, due to unavailability of experimental estimates of the scattering times. Yet, we can make some qualitative simulations. The disorder related to C substitution mainly affects the σ bands, as these bands are spatially localized in the B planes, where the C substitution occurs. Experimental data of C substituted MgB$_2$ samples confirm this view [51,52]. We thus reasonably assume that, with increasing C doping, the scattering time $\tau_{0\sigma}$ rapidly decreases, the parameter $\eta=\tau_{0\pi}/\tau_{0\sigma}$ becomes much larger than unity and S$_{diff}$≈S$_{diff\_\pi}$. Moreover, electrons are filled in the system upon C substitution on the B site. In figure 12 the so calculated S$_{diff}$ is displayed. The quantitative agreement with experiments is poor, as the magnitude of the calculated S$_{diff}$ is much lower than the measured one, mainly at low doping levels, however the calculated data reproduce the sign change of S with increasing doping for x between 0.05 and 0.075. Hence the comparison of measurements and calculations indicates that the suppression of S magnitude at low doping and the change in S sign at high doping (see figure 11b) are explained in terms of electron filling in the π band (see figure 6b).

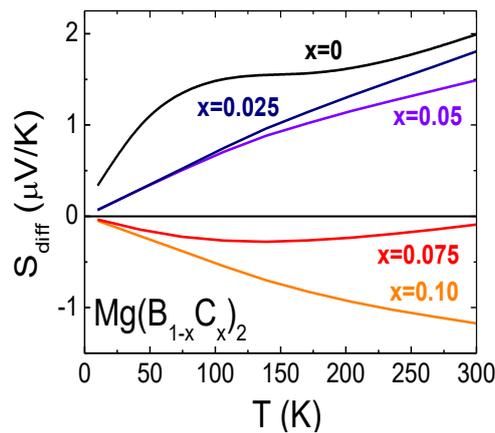

**Figure 12:** Diffusive contribution to thermopower calculated for increasing electron doping and increasing $\tau_{0\pi}$ scattering rate, as it is the case of increasingly C doped samples. The corresponding experimental S(T) curves are displayed in figure 11b.

## 6. Discussion

We first consider the series of irradiated samples, where the effect of increasing disorder certainly plays the major role. The evident difference between the experimental S(T) curves (figure 7) and the calculated diffusive contributions (figure 8) in the pristine and less irradiated samples suggests a sizeable contribution of the phonon drag term in these samples. Yet such difference tends to vanish with increasing neutron irradiation and eventually for the most irradiated sample the experimental curve and the calculated $S_{diff}$ are comparable in magnitude and shape, with an almost linear temperature dependence. This indicates that the discrepancy at low disorder is related to the presence of a phonon drag term, which is sensitive to disorder and thus easily suppressed in dirty samples. The same behavior is observed in Al doped samples, for which evident differences between the experimental S(T) curves (figure 9) and the calculated diffusive contributions (figure 10) are found, particularly for the undoped and low doped samples. In order to support this scenario, we plot in Figure 13 the excess Seebeck coefficient, defined as the difference between the measured S and the calculated diffusive $S_{diff}$, $S_{ex}=S-S_{diff}$, as a function of the residual resistivity $\rho_0$, which gives an estimate of the level of intrinsic disorder in the samples. We analyze value of $S_{ex}$ at T=200 K because near this temperature it is expected that the phonon drag has its maximum value. We do not consider the C-doped samples, for which a quantitative evaluation of $S_{diff}$ is lacking. $S_{ex}$(T=200K) versus $\rho_0$ follows a universal trend, despite the fact that we are comparing samples disordered with very different methods. It decreases monotonically with increasing $\rho_0$ from ~8μV/K in the pure sample to negligible values for the most disordered samples. This gives support to the idea that $S_{ex}$ is actually due to the phonon drag contribution. We observe that in the undoped clean sample the phonon drag term gives the dominant contribution to the measured Seebeck coefficient as compared to the diffusive term and also in low disorder sample it contributes substantially.

The sizable contribution of the phonon drag term in S and its sensitivity to disorder, and consequently also its sample-dependent character, explains as well the scattering of experimental data collected from literature of room temperature Seebeck coefficient of undoped $MgB_2$ (see Table I).

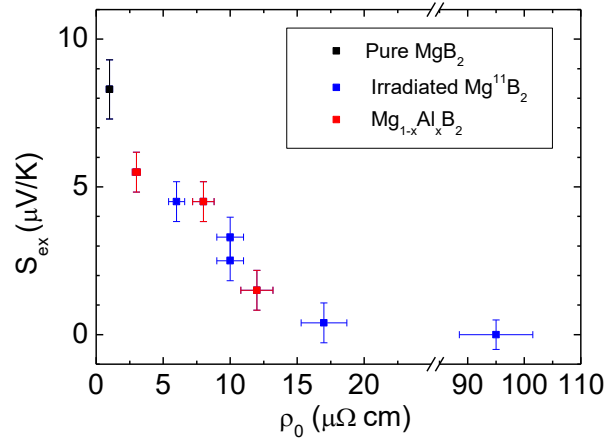

**Figure 13:** Excess Seebeck calculated as the difference between the measured S and the calculated diffusive S, $S_{ex}=S-S_{diff}$, at 200K, plotted as a function of the residual resistivity $\rho_0$. Error bars are estimated from experimental uncertainty on the measured S and $\rho_0$.

In Al doped samples the effect of electron band filling on the diffusive S is masked by the sizeable phonon drag S, however in the C doped series of samples the effect of band filling is visible, as evidenced by the sign change of S. This difference can be ascribed to the different disorder induced by C and Al substitutions. Al substitution does not affect σ-bands and the system remains in the clean limit, as evidenced from upper critical field measurements [53]. In this case, the phonon-drag mechanism, which couples mainly phonons mainly with σ-bands, is not appreciably reduced.

On the contrary, C substitution affects mainly the σ-bands and the Seebeck coefficient gives a two-fold evidence of this effect: i) reduction in magnitude and change of sign with increasing substitution, as a consequence of crossing from σ to π dominated $S_{diff}$ (see figure 6); ii) almost linear temperature dependence with increasing C doping, due to fast suppression of the phonon-drag term and crossover from the clean to the dirty limit of $S_{diff}$ (see figure 3). The arguments about tendency to S(T) linearity with increasing disorder also apply to the case of neutron irradiated samples.

Our interpretative scenario also accounts for literature results on doped $MgB_2$ samples. In the case of isoelectronic substitutions on the Mg site, either nonmagnetic with Fe [28] (which has negligible effect on $T_c$) or magnetic with Mn [22] (which suppresses $T_c$ severely), the π bands are progressively affected by disorder and η decreases, driving transport to a σ dominated regime. As a result, $S_{diff}$ increases in magnitude (see figure 2). At the same time, the suppression of the phonon drag term $S_{drag}$ by disorder is mild, as the σ band is only marginally affected by disorder, therefore the S curves maintain the broad bump above 150K, characteristic of the undoped clean samples. These multiple effects are also present in our Al doped samples, where Al is substituted on the Mg site, yet in this case there is the additional effect of electron charge effectively introduced in the bands. As for Co addition [34], its effect seems mainly that of introducing disorder in the σ band and suppressing $S_{drag}$, which explains the observed decrease in magnitude and progressive tendency to linear temperature dependence of S. Likewise, in the case of Be substitution on the B site, $MgB_{2-x}Be_x$, disorder introduced in the σ band may be largely responsible for the observed decrease in S magnitude [25], although in this case also hole doping in the σ band has certainly an effect on S(T), determining the sign change at low temperature for x=0.6.

## 7. Conclusions

In this work, by using a combined theoretical and experimental approach, we investigate the roles of selective disorder and selective doping in tuning the relative contributions of σ and π bands to normal state thermoelectric transport. We carry out calculations of the diffusive contribution to the Seebeck effect based on ab initio band structure calculations and solution of the Boltzmann equation in the relaxation time approximation. From the calculated curves, we show that the diffusive S(T) can be tuned in magnitude, shape and sign by controlling relative disorder and electron filling in the two bands. We measure the Seebeck effect in neutron-irradiated $Mg^{11}B_2$ and $Mg_{1-x}Al_xB_2$ and $Mg(B_{1-x}C_x)_2$ doped series of samples and compare our experimental results with simulations of the diffusive S(T) for the very same samples. In irradiated samples we observe a progressive decrease in S magnitude and tendency to linear S(T) temperature dependence with increasing irradiation fluence, which we explain in terms of suppression of the phonon drag term and crossover of $S_{diff}$ from the clean to the dirty limit. In C and Al doped samples, we observe a different behavior, despite in both cases electron are filled in the system. In Al doped samples, S is hardly affected in magnitude, whereas in C doped samples S is suppressed and becomes more linear in temperature, eventually changing sign with increasing substitution. We interpret these findings in terms of different kinds of disorder, affecting either σ or π bands, which results in sizeable tuning of both the diffusive and the phonon drag terms.

This work evidences that the close comparison of theoretical and experimental analyses of Seebeck effect in $MgB_2$ is successful in gaining information on scattering processes and multiband transport.